%% file: M.tex
 \documentclass[pra,floatfix,superscriptaddress,nofootinbib,twocolumn]{revtex4}
\usepackage{dcolumn,amsmath}
\usepackage{graphicx}
\usepackage{bm}
\usepackage{amssymb}
\usepackage{hyperref}

\usepackage{xcolor}
\usepackage{pdfcomment}

\DeclareMathOperator{\Tr}{Tr}
 \newcommand{\sect}[1]{section~\ref{#1}}
 \newcommand{\Table}[1]{Table~\ref{#1}}
 
 \newcommand{\Eq}[1]{Eq.~(\ref{#1})}
 \newcommand{\Eqs}[1]{Eqs.~(\ref{#1})}
 \newcommand{\etal}{{\it et~al.}\ }
 \newcommand{\abinitio}{{\it ab~initio} }
 
 \newcommand{\appx}[1]{Appendix~\ref{#1}}
\begin{document}
\title{Method of evaluating chemical shifts of X-ray emission lines in molecules and solids}
\author{Yuriy V. Lomachuk}\email{jeral2007@gmail.com}
\author{Anatoly V. Titov} \email{anatoly.titov@gmail.com}
\homepage{http://www.qchem.pnpi.spb.ru}
 \affiliation{B.P.\ Konstantinov Petersburg Nuclear Physics
              Institute, Gatchina, Leningrad district 188300, Russia}
\affiliation{Department of Physics, Saint Petersburg State University, Saint Petersburg, Petrodvoretz 198504, Russia}

  \date{\today}

\begin{abstract}
        A method of evaluating chemical shifts of x-ray emission lines for period four and heavier elements is developed. This method is based on the relativistic pseudopotential model and one-center restoration approach [Int. J. Quantum Chem.\ {\bf 104}, 223 (2005)] to recover a proper electronic structure in heavy-atom cores after the pseudopotential simulation of chemical compounds. The approximations of instantaneous transition and frozen core are presently applied to derive an expression for chemical shift as a difference between mean values of certain effective operator.  The method allows one to avoid evaluation of small quantities (chemical shifts $\sim0.01{-}1$~eV) as differences of very large values (transition energies $\sim 1{-}100$~keV in various compounds). The results of our calculations of chemical shifts for the $K{\alpha_1}$, $K{\alpha_2}$, and $L$ transitions of group-14 metal cations with respect to neutral atoms are presented. Calculations of $K{\alpha_1}$-line chemical shifts for the Pb core transitions in PbO and PbF$_2$ with respect to those in the Pb atom are also performed and discussed. The accuracy of approximations used is estimated and quality of the calculations is analyzed.
\end{abstract}

\maketitle
\input{intro}
\input{sec1}

\input{sec2}
\input{sec3}
\input{sec4}
\input{conc}
\appendix
\input{app_a}
\input{app_b}
\input{app_c}

\bibliography{./bib/JournAbbr,./bib/Lomachuk,./bib/TitovLib,./bib/QCPNPI}
\end{document}

%% file: intro.tex

\section*{Introduction}

One of the most efficient methods of experimental study of the electronic density distribution on atoms in materials is analysis of chemical shifts of the x-ray emission spectra (XES) \cite{Siegbahn:08}. This method is based on the fact that the radiation caused by an electronic transition between atomic core shells depends on the redistribution of densities of the valence and outermost core electrons from one compound to another. The typical order of magnitude of XES energies is  $1{-}100$~keV for one-electron transitions in atoms with $Z{\gtrsim}30$; such electrons are designated below as transition core (TC) electrons, emphasizing that the electronic states involved in the transitions are spatially localized in the atomic core regions.

The XES chemical shift is the difference between energies of the same characteristic x-ray line in different compounds; to describe the redistribution of electronic densities based on experimentally observed chemical shifts, which usually range \hbox{within~$\sim0.01{-}1$~eV}, one needs to compare the experimental data with theoretical calculations of different atomic configurations. This comparison allows one to determine the occupation numbers of $d_{3/2}$ and $d_{5/2}$ shells in transition metals and those of $f$ shells in lanthanides and actinides \cite{Sumbaev:78}. These shells are spatially localized in the atomic core region, while they can be rather considered as valence from an energetic point of view. As a result, the partial charges of $d$ and $f$ elements in different compounds can be studied and some information about the electronic spin densities of these shells can also be extracted. However, the occupation numbers of outermost valence ($s$ and $p$) shells cannot be identified unambiguously. Nevertheless, taking more x-ray transitions into consideration, a clearer picture about the state (electron configuration) of a given atom in a compound can be extracted, meaning the outermost core- and valence-electron densities.

The well-known theoretical studies of chemical shifts are usually based on analyzing of isolated atoms. A chemical shift is then represented as a function of occupation numbers of valence shells and the aim of these studies is reduced to calculating of these occupation numbers to better reproduce the experimental chemical shift values.

 Below we give a short review of papers devoted to the chemical shift theory during the last decade and related to the subject of our paper. 

Raj \etal \cite{Raj:98} showed that it is possible to obtain the $3d$ electron population in a transition metal in various compounds by comparing the experimental data of K$\beta$-to-K$\alpha$ x-ray intensity ratios with results of the atomic multiconfiguration Dirac--Fock computations (see Refs. \cite{Raj:02,Pawlowski:02} and references therein). It is shown in Ref.~\cite{Raj:02} that the experimentally determined valence electronic structures for all the metals except V, Cr, and Mn agree reasonably well with the results of augmented plane-wave band structure calculations.

In 2004, Batrakov \etal \cite{Batrakov:04} studied the influence of relativistic effects on the chemical-shift values of XES in compounds of uranium. The authors considered the chemical shift as a sum of two values, the first one being the chemical shift of the centroid energy of the x-ray line (which is an average of the chemical shifts over the corresponding multiplet), while the second one is a correction to the chemical shift due to the spin-orbit interaction. These terms are represented as functions of occupation numbers (or ``charges'') of valence shells. The experimental data on the chemical shift of uranium $L$ lines in UF$_4$ and UO$_3$ with respect to UO$_2$ as the reference are tabulated. Interpreting these data with the help of atomic Dirac--Fock-based analysis, the changes in occupation numbers of the $5f$ and $6d$ shells of uranium in the compounds above are determined with respect to those in UO$_2$. On the basis of these calculations, the authors conclude that the relativistic correction to the total value of the chemical shift is independent of the oxidation state of uranium and is determined by an intra-atomic redistribution of the electron density between the subshells $5f_{5/2}$ and $5f_{7/2}$. Thus, it is shown that the relativistic component of XES chemical shift allows one to determine the distribution of $f$-electrons within subshells.

The form of the $K$ x-ray line of medium and heavy atoms was studied by Polasik \etal \cite{polasik:06} in 2006. Based on the multiconfiguration Dirac--Fock calculations with quantum electrodynamic corrections, the $K$ line and its satellites on Pd, Tb, Ta, and Th with and without the natural line widths are modeled.
The theoretical decomposition of the spectrum allows one to predict the overlap and the resolution of different groups of lines. The authors study effects of removing the electrons from $3p$ and $3d$ shells on the structure and shapes of x-ray spectra. The x-ray study of the $K$-line form allows one to perform reliable quantitative analysis of the experimental data.


To attain  a qualitative agreement between \abinitio computations of small XES chemical shifts in molecules (and, especially, in materials) and experimental data, the molecular calculation must be performed with an accuracy that is usually a challenge in practice to modern computational methods. The transition energy is ordinarily calculated as a difference between total energies of two low-lying many-electron states of the system containing a given atom. In turn, the chemical shift is a difference between the transition energies in the  studied system and a reference system. Therefore, when using ``direct'' computational procedures, the small chemical-shift value is obtained as a ``double-difference'' of large energy eigenvalues and its magnitude often lies within the error margin of the calculation.

The papers discussed below are devoted to evaluating of binding energies and ionization potentials of the core electrons in different compounds.

Takahata and Chong (2003) \cite{Takahata:03} analyzed the problem of computing the binding energies of the atomic core electrons in light-atom molecules within the framework of density functional theory (DFT). The authors calculated the binding energies as differences between the total molecular energies of the Kohn--Sham solutions for the ground state and states with a core hole. It is shown that the accuracy of the calculation strongly depends on the chosen density functional and basis set. For various combinations of the functionals and basis sets, 59 binding energies of the core electrons are determined. For the most accurate studies, the average absolute deviation from the experimental values is 0.16~eV.
Segala and Chong (2010) \cite{segala:10} calculated the ionization energies of the $1s$ electron of sulfur or phosphorus in different compounds using DFT. The ionization energies were calculated as differences between total energies of the corresponding states. The authors used various density functionals in their studies. The deviations from the experimental data are within 0.5~eV. The authors also analyzed how the hybridization of orbitals affects the ionization energy of the $1s$ state.
It should be emphasized, however, that the quality of evaluated energies within DFT
significantly depends on applied functionals. Moreover, there are some problems in choosing the appropriate functional for certain heavy atoms and  systems (e.g., see Ref. \cite{Zaitsevskii:13b} and references therein), whereas for systems containing light atoms, one can usually estimate the errors associated with the density functional approximations by comparing the DFT results to those obtained within \abinitio approaches.

The method of computing the core electron energies based on second order M{\"o}ller--Plesset perturbation theory was discussed by Shim with colleagues (2011) \cite{shim:11}. This method was applied to obtain the binding energies of core electrons of the C, N, O, and F atoms. The key feature of this method is in using the mixed basis set which consists of all-electron basis functions for the considered atom (with ionized core state) and the reduced basis sets for all the other atoms of the compound to be applied together with the pseudopotentials for the atoms. The authors show that the accuracy of developed method is about 0.16~eV.

Holme \etal (2011) \cite{holme:11} evaluated chemical shifts of the ionization energies of the $1s$ electron in carbon for a variety of organic compounds with errors of no more than 20 meV for chemical shifts and no more than 30 meV for ionization potentials. This level of accuracy comes close to that of modern experiments. Chemical shifts are calculated using various methods: Hartree--Fock, M{\"o}ller--Plesset perturbation theory, coupled clusters, and DFT.  The authors show that the errors for determining the chemical-shift values by DFT are about three times larger than those from the coupled cluster theory.

In 2010, Lee \etal \cite{lee:10} presented a systematic study of 12 ferric and ferrous $K_{\beta}$ lines in different compounds. The factors contributing to the shift of the main line of the spectra and its satellites are studied both experimentally and theoretically. It was shown that the shift of the main line depends mainly on the spin state of Fe, while the valence-to-core region of the spectra (with the electronic transitions from valence shells to core) have greater sensitivity to changes in the chemical environment. DFT is used to calculate transition energies and intensities at the one-electron-approximation level. The authors estimate the errors of the methods used in the studies within a few tenths of eV. It is shown that the electric dipole transitions from the $np$ to $1s$ shells of iron dominate in the spectra.

DeBeer and Neese (2010) \cite{debeer:10} proposed a method of evaluating the x-ray absorption spectra based on DFT. This method requires a preliminary calibration based on experimental data.  Contributions to the transition energies due to the scalar relativistic effects are taken into account. The authors estimate the errors in calculating of transition energies from the method at the level of magnitude of 0.1~eV.
Lancaster \etal (2011) \cite{lancaster:11} studied the form of $K{\beta}$ lines for neutral and singly ionized ferrocene. The claim that XES provides information about molecular orbital energies is justified by studying the valence-to-core regions of these spectra. The DFT calculations show that the valence-to-core electronic transitions occur due to admixture of Fe $4p$ orbitals to the valence orbitals of the considered compounds.

%
\input{tabl0}
To summarize, the computational errors for energies of x-ray emission and absorption spectra, presented in papers discussed above \cite{Takahata:03, segala:10, shim:11, holme:11, lee:10, debeer:10, lancaster:11} and listed in \Table{tabl0}, are at 0.1~eV by the order of magnitude. The accuracy of the methods applied in these studies is generally sufficient for evaluating the chemical shifts of x-ray lines for light atoms or low-energy x-ray transitions (typical chemical shift values are within \mbox{$0.5{-}1$~eV}). However, in the case of the x-ray transitions between inner core shells of heavy atoms, the errors of evaluating the corresponding energies are in general notably greater than the chemical shift values for the transitions. This problem is aggravated by the computational complexity of the relativistic calculations required for studying heavy-atom systems.
We illustrate this in \sect{sec4n} taking the inner-core transitions in lead as an example.

Our paper is devoted to the development of an \abinitio\ approach to calculate XES chemical shifts in heavy-atom compounds. The approach is based on precise relativistic pseudopotential models \cite{Titov:99, Mosyagin:06amin, Mosyagin:10a} and one-center restoration techniques \cite{Titov:05a, Titov:06amin}
   to provide an optimal combination of computational savings and high accuracy simultaneously.

%% file: tabl0.tex
\begin{table}[h]
\centering
	\caption{Typical errors in \abinitio calculations of ionization potentials, core-electron binding energies, and x-ray chemical shifts.}
\scalebox{0.8}{%
\begin{tabular}{lc}
\toprule
Computational details 			&  Average error, eV\\
\hline
Takahata and Chong (2003), DFT/PW86-PW91\footnote[1]{\,DFT study of core-electron binding energies of C to F elements with exchange-correlation functional PW86-PW91 \cite{Takahata:03}.}
          &   $0.16$   \\
Segala and Chong (2010), DFT\footnote[2]{\,DFT study of 1$s$ ionization energies for P- and S-containing molecules with exchange-correlation functional becke00xx(xc) \cite{segala:10}.} &   $0.2$\\
Lee \etal (2010), DFT\footnote[3]{\,DFT study of the ferric and ferrous $K{\beta}$ line energies with exchange- correlation functional BP86 \cite{lee:10,lancaster:11}.}         &   $0.1$ $-$ $0.5$       \\
DeBeer and Neese (2010), DFT\footnote[4]{\,DFT study of  sulfur $K$-edge x-ray absorption transition energies \cite{debeer:10} with a range of contemporary functionals.}      &   $0.1$         \\
Shim \etal (2011), MBPT\footnote[5]{\,MBPT study of 1$s$ core-electron binding energies for the C, N, O, F atoms in molecules  \cite{shim:11}.} &   $0.163$        \\
Holme \etal (2011), HF, MBPT, CC, DFT\footnote[6]{\,Study of 1$s$ carbon energies for a variety of organic compounds using various methods: Hartree--Fock, M{\"o}ller--Plesset perturbation theory, coupled clusters,  and DFT \cite{holme:11}.}  &   $0.03$(CC) $-$ $0.1$(DFT)    \\
\toprule
\end{tabular}%
}
\label{tabl0}
\end{table}

%% file: sec1.tex
     \section{Features of chemical shift theory in heavy-atom systems}
 \label{sec1}
 \label{sec1n}

The precise \abinitio study of heavy-atom compounds is a complicated problem from the technical point of view due to the importance of accounting for relativistic and correlation effects simultaneously.  Below we consider the one-electron $K, L, M$ transitions in bound and free atoms starting from period-four  elements, i.e.,  the transitions of interest take place between the core shells. For the electrons occupying these shells, the relativistic effects are important. Straightforward relativistic treatment of such properties requires applying four-component techniques for all electrons in a compound,  and not just for the core region of a given atom where the electronic transition takes place. The total number of electrons can be very large in polyatomic systems (solids, clusters, supramolecular structures etc.) which are of primary interest in practice, and this brute force way of evaluating chemical shifts is extremely consuming.
However, the specifics of the problem under consideration allows one to introduce a number of sufficiently valid approximations, considered below, and to avoid the use of an all-electron relativistic treatment when studying heavy-atom systems. The proposed approximations allowing one to reduce dramatically the efforts are based on a natural supposition that one may divide the set of one-electron states of the system into the following groups taking into account their role in the considered inner-core electron transitions:

\begin{enumerate}
  \item[Ic:]

The group of the states, which are localized in a small inner core (Ic) region of a given atom; their wavefunctions (described by four-component one-electron spinors) are nearly the same in different compounds containing this atom. One can treat these states as ``frozen'' with an accuracy sufficient for applications (see \sect{sec3n}); correlation effects for these states can also be neglected. We will consider below the x-ray transitions between the shells belonging to this group only.

  \item[Oc:] 

The group of states belonging to the outer core (Oc) region, which are relaxed only slightly in a given atom chemically bound
 in one compound against the other,
but the energy contributions to the chemical shifts from their relaxation can be of the same order of magnitude as those from the valence shells (see below). Nevertheless, one may take account of small differences between the corresponding wave functions of the atomic Oc states in distinct compounds by using the lowest orders of perturbation theory.

  \item[V:]

The group of valence states (V). We assign to this group either all the valence states of the system (rather for few-atom molecules), or only those valence states (bonding or antibonding orbitals, etc.) which have notable wavefunction amplitudes in the valence area near a given atom (in polyatomic compounds).%
\footnote{%
The total number of the valence states in the case of rather complex systems, solids etc.\ can be too large to be treated explicitly for the chemical shift's evaluation. After some electronic structure calculation of a complex system, the valence states are usually represented as combinations of either localized (Gaussian) basis functions or plane waves. In these cases all these valence states may be reexpanded in a spherical region around the given atom on partial waves. The radius of the sphere must be greater than the radius, where one-electrons states are smoothed within the pseudopotential treatment \cite{Titov:99}, but lesser than the distance to the nearest atom. For each valence state in this reexpansion only the terms which significantly contribute to the chemical shift (see next sections) should be saved. Thus, the computational complexity is reduced due to minimizing the basis-set size used in calculating system.%
} 
The occupation numbers and space distribution of these states can differ significantly for various compounds. These states usually form or contribute notably to the chemical bonds of the  considered atom with its neighbors. Some of basis functions which are most important to take into account for the correlation of the Oc and V electrons can also be assigned to this group.

  \item[W:]

We denote the combined group of the states belonging to either V or Oc subspaces by the symbol W, ${\rm W=Oc\bigcup V}$. We will use this designation when the distinction between the states from groups V and Oc is not important.
%

  \item[R:]

All other one-electron states, which are not attached to one of the former groups, are assigned to group R (rest); in particular, core states of other atoms in the system belong to this group.
It is shown below that the influence of these states on the properties of our interest are not essential (since their densities in the inner core of a selected atom are negligible) and the states R can be excluded from consideration concerning chemical-shift theory.
%
%
\end{enumerate}

One-electron states may be classified in such a way already after preliminary self-consistent-field (SCF) treatment of low-lying electronic states of a given atom, thus avoiding calculation of the whole system of interest. The given classification allows us to take into account a number of features of the problem under consideration to construct a robust model describing chemical shifts. This approach establishes a direct link between observable chemical shifts and the corresponding quantum-mechanical expectation values and leads to serious savings at the computational stage:
\begin{itemize}
   \item 

Typical times of transitions between the Ic shells are $\tau \sim 10^{-16}\div 10^{-13}\, \hbox{s}$ (for heavy atoms), the times of transitions of Oc and V electrons are \hbox{$\tau' \sim 10^{-12}\div 10^{-8}\, \hbox{s}$} \cite{Labzowsky:96}; thus, the V and Oc shells do not change significantly during the Ic electron transitions.
In the present study, the many-electron effects such an Auger transitions of Ic electrons or radiative transitions of V and Oc electrons are not considered. Effects of the relaxation of the system during the
  fast Ic transition are usually small and are not taken into account here.%
 \footnote{
It can be shown that the latter corresponds to the approximation in which all the transitions between the states belonging to some fixed initial and final shells, $I$ and $F$, have equal probabilities and the energies differ from each other by the values which are much less than the linewidth. For the case of our interest typical values of these linewidths are $\Gamma \sim 2 \div 60$~eV.}
These constraints allow us to compute transition energies as differences of ionization potentials from the final and initial shells. Note, however, that the given constraints are used rather for the manifestation of our model and are not mandatory in general when some electron relaxation effects are taken into account.

     \item

One may very accurately take into account the relativistic effects for valence and outer-core electrons in heavy-atom compounds by using the pseudopotential approach \cite{Mosyagin:10a}. Contributions to the energy of the considered Ic transition in a given atom from interaction of TC electrons with those occupying the group-R states and nuclei of other atoms largely compensate each other, therefore, they can be neglected for the considered problem with good accuracy (see Appendix~B for details).  
\end{itemize}

%% file: sec2.tex
\section{Theory of chemical shifts for X-ray emission spectra}
\label{sec2}

Denote the many-electron wave functions of initial and final states, which are obtained after electron ejection from some Ic shell and after transition of the other electron to the vacant Ic state (accompanied by x-ray emission), as $\left | \Psi_i \right \rangle$ and $\left | \Psi_f \right \rangle$, correspondingly. Since we can usually neglect the correlation effects for the Ic electrons, these shells are well described within the Dirac--Fock model.

We define a parent state of the system under consideration with completely occupied Ic shells, in which the system was before electron ejection, as the ground eigenstate $\left | \Psi_0 \right \rangle$ of some appropriate Hamiltonian $H_0$ describing our system:

\begin{equation}
        H_0 \left | \Psi_0 \right \rangle = E_0\left | \Psi_0 \right \rangle\ .
            \label{sec2:eq2}
\end{equation}
In the sudden-transition approximation, the many-particle states $\left | \Psi_i \right \rangle$ and $\left | \Psi_f \right \rangle$ can be obtained from the parent state by removing an electron occupying the one-particle $\left | i \right \rangle$, and $\left | f \right \rangle$ states belonging to the Ic shells $I$ and $F$, respectively, whereas all the other one-particle states of the system are considered unchanged.
Below we use the frozen-inner-core approximation, i.e.\ neglect the effects of correlation and relaxation of the inner core one-electron states (see the next section for details).
Let us write the Hamiltonian $H_0$ in the second-quantization representation:

\begin{equation}
        H_0 = \sum\limits_{pq}  h_{pq}a^{+}_pa_q +
        \frac{1}{2}\sum\limits_{pqrs}V_{pqrs}a^{+}_pa^{+}_qa_ra_s\ ,
        \label{sec2:eq6}
\end{equation}
\begin{equation}
	h = T+ V^{\rm{A}} + \sum\limits_{\rm{A'} \ne \rm{A}} V^{\rm{A'}}\ .        
        \label{sec2:eq9}
\end{equation}
In this expression the one-electron operator $h$ includes the kinetic energy of electrons, their interaction with the nucleus of atom $A$, in the core of which the transition occurs, and the interaction with nuclei of the other atoms, $A'$, in the system. The two-electron operator, $V=1/r_{12}$, takes into account the Coulomb interaction between electrons
(in general, one can easily include the relativistic interactions between electrons as well, see Ref. \cite{Mosyagin:06amin}).

In the sudden-transition approximation, the final and initial states, $\left | \Psi_f \right \rangle$  and $\left | \Psi_i \right \rangle$, can be written as

\begin{equation}
        \begin{split}
                \left | \Psi_f \right \rangle =  a_f \left | \Psi_0 \right \rangle\ ,\\
                \left | \Psi_i \right \rangle =  a_i \left | \Psi_0 \right \rangle\ . 
        \end{split}
        \label{sec2:eqs10}
\end{equation}
In the framework of the frozen-core approximation, the energies of these states, $E_x (x=f,i)$, are

\begin{equation}
        E_x = E_0 - h_{xx} - \langle \Psi_0 |\sum\limits_{rs}(V_{xxrs} - V_{xrxs})a^{+}_ra_s| \Psi_0 \rangle\ ,
        \label{sec2:eq7}
\end{equation}
where the summation indices $r,s$ and the indices of the transition core states do not overlap.

{\tolerance=1000
Using the one-electron density matrix 
\hbox{$\bm{\rho}_{rs} = \left \langle \Psi_0 | a^{+}_r a_s | \Psi_0
\right\rangle$}, write $E_x$ as

\begin{equation}
                E_x = E_0-h_{xx} - \Tr[\bm{F}_{xx}\bm{\rho}]\ ,
      \label{sec2:eq8}
\end{equation}
where $F_{xx}^{ab} = V_{xxab} - V_{xaxb}$.

Let us use the projectors on the introduced above subspaces W, R, and Ic:
$P_{\mathrm{W}}$, $P_{\mathrm{R}}$, and $P_{\mathrm{Ic}}$.
Acting on $\bm{\rho}$ by the projectors $P_{\mathrm{Ic}} + P_{\mathrm{W}} + P_{\mathrm{R}} = 1$ from the left- and right- hand sides we obtain

\begin{equation}
        \begin{split}
                \bm{\rho} = \bm{\rho}_\mathrm{Ic} + \bm{\rho}_\mathrm{W}+\bm{\rho}_\mathrm{R} + \bm{\rho}_\mathrm{WR} + \bm{\rho}_\mathrm{RW},\\
        \bm{\rho}_\mathrm{Ic}= P_{\mathrm{Ic}} \bm{\rho}  P_{\mathrm{Ic}},\\
        \bm{\rho}_\mathrm{W} =  P_{\mathrm{W}} \bm{\rho}  P_{\mathrm{W}},\\
                \bm{\rho}_\mathrm{R}=  P_{\mathrm{R}} \bm{\rho}  P_{\mathrm{R}},\\
                \bm{\rho}_\mathrm{WR} =  P_{\mathrm{W}}\bm{\rho} P_{\mathrm{R}}\, ,
  \bm{\rho}_\mathrm{RW} = P_{\mathrm{R}}\bm{\rho} P_{\mathrm{W}}.
        \end{split}
        \label{sec2:rhodef}
\end{equation}  

\begin{sloppypar}
In the above expression the off-diagonal terms $ P_{\mathrm{Ic}} \bm{\rho}  P_{\mathrm{W}}$, $ P_{\mathrm{Ic}}\bm{\rho} P_{\mathrm{R}}$, and $ P_{\mathrm{W}}\bm{\rho}P_{\mathrm{R}}$, as well as their Hermitian conjugates, vanish because of the frozen core approximation used.

\end{sloppypar}
Substituting this expression for the density matrix into~(\ref{sec2:eq8}), we obtain

\begin{equation}
        E_x = E_0 - E^{\rm Ic}_x - \varepsilon_x\ , 
        \label{sec2:eq223}
\end{equation}
where

\begin{equation*}
        \begin{split}
        E^{\rm Ic}_x = T_{xx} + V^{\rm A}_{xx}+ \Tr[\bm{F}_{xx}\bm{\rho}_\mathrm{Ic}], \\
        \varepsilon_x = \sum\limits_{\rm A' \ne A} V^{\rm A'}_{xx} +
        \Tr[\bm{F}_{xx}(\bm{\rho}_\mathrm{W} +\bm{\rho}_\mathrm{RW}+\bm{\rho}_\mathrm{WR}+\bm{\rho}_\mathrm{R})]\ .
        \end{split}
    \end{equation*}

The one-electron state $\left | x \right \rangle$ and corresponding energy $\varepsilon_x$ are eigenvector and eigenvalue of some effective one-electron operator $h^{\rm eff}$ defined on the subspace $X$ of one-electron Ic states. The matrix elements of this operator are

\begin{equation}
        h^{\rm eff}_{pq} = \sum\limits_{\rm A'} {V^{\rm A'}_{pq}} +
        \Tr[\bm{\bm{F}}_{pq}(\bm{\rho}_\mathrm{W} +\bm{\rho}_\mathrm{RW} + \bm{\rho}_\mathrm{WR}+\bm{\rho}_\mathrm{R}) ],\ p,q\in X\ .
        \label{sec2:eq10}
\end{equation}
.
Let us mark out a spherical area around the atom where $\left | f \right \rangle$ and $\left | i \right \rangle$ are localized. The radius of this area, $R_c$, is such that the amplitudes of the Ic states are negligible outside the area, whereas the amplitudes of the R states are negligible inside. Denote a submatrix $\bm{\rho}_\mathrm{W}$ of the density matrix which corresponds to the electron distribution inside the introduced spherical area, $\bm{\rho}_\mathrm{W}({\bf r,r'}): |{\bf r}|,|{\bf r'}| < R_c$ (see Appendix~A), as $D$. Then we can rewrite expression~(\ref{sec2:eq10}) as

\begin{equation}
        h^{\rm eff}_{pq} =  \Tr[\bm{\bm{F}}_{pq}D] + V^{\rm{ext}}_{pq}\ .
        \label{sec2:eq8mod}
\end{equation}
The operator $V^{\rm{ext}}$ describes the interaction of the TC electron with atomic nuclei and electrons outside the sphere with radius $R_c$. For $r<R_c$ we can represent this operator as local (see Appendix~A):

\begin{equation}
        V^{\rm{ext}} \approx \sum_{km} U_{km}r^k Y_{km}(\Omega),\mbox{ for
        $r<R_c$ } .
        \label{sec2:eq12}
\end{equation}

Let us consider transition energies $\Delta E_{fi}$ and $\Delta E_{f'i'}$, where states $\left | f \right \rangle$, $\left | f' \right \rangle$ belong to the $F$ shell and $\left | i \right \rangle$, $\left | i' \right \rangle$ belong to the $I$ shell.  In the case of interest these energies differ much less than the natural line widths.\ %
\footnote{
Energies $\varepsilon_x$ and $\varepsilon_{x'}$ (corresponding to states $\left | x \right \rangle$ and $\left | x' \right \rangle$ from the same shell)
   do not coincide with each other in general
because of spherically asymmetric contributions from the interaction of the TC electron with the valence electrons and the other atoms of the system. Typical values of their differences are less than 10 meV (see Appendix~B). This is much less than the Ic transition linewidths (typical values for which are 2 eV by the order of magnitude for sufficiently heavy atoms).
} 

Thus the experimentally observed transition energy is practically equal to the average over all the transition energies between the states from shells $F$ and $I$: 

\begin{equation}
        \overline{\Delta E_{FI}} =\frac{1}{N}\sum\limits_{f \in {F} , i \in
        {I}}\Delta
        E_{fi}\zeta_{fi}\ , 
        \label{eq:sdfsdf}
\end{equation}
where $N = \sum\limits_{f \in F,i\in I} \zeta_{fi}$, and $\zeta_{fi}$ are the relative probabilities of transitions between one-electron states $\left | f \right \rangle$ and $\left | i \right \rangle$.

Let us use the average relativistic configuration approximation and consider the probabilities of all the transitions from $F$ to $I$ to be equal to each other. Then $\zeta_{fi}=1$, $N=(2j_F+1)(2j_I+1)$ and the average transition energy is 
\footnote{ 
The values of \unexpanded{$E^{\rm Ic}_{x}$, $x=i{,}\,f$} are contributions to the transition energy from the kinetic energy of the TC electron and from the interaction of the TC electron with other (frozen) inner-core electrons, and the nucleus of the considered atom. Due to the spherical-symmetry approximation used for these states, \unexpanded{$E^{\rm Ic}_{x}=E^{\rm Ic}_{x'}=E^{\rm Ic}_{X}$}.
}

\begin{equation}
	\begin{array}{r}
        \overline{\Delta E_{FI}} \approx  E^{\rm Ic}_{F}-E^{\rm Ic}_{I} +\\
+\frac{1}{N}\left(\Tr\left[\left\langle f |h^{\rm eff}| f' \right\rangle\right]-
        \Tr\left[\left\langle i |h^{\rm eff}| i' \right\rangle\right]\right).
	\end{array}
        \label{eq:deltaEshfi}
\end{equation}

The expression above differs from the exact average energy due to the inequality of the probabilities of the various transitions from $F$ to $I$. The value of this difference is mainly determined by spherically asymmetric contributions in $h^{\rm eff}$.

Within the framework of the relativistic average configuration approximation, the expression for
$\overline{\Delta E_{FI}}$ can be written as

\begin{equation}
        \overline{\Delta E_{FI}} = E^{\rm Ic}_{F} - E^{\rm Ic}_{I} +
        \Tr[\overline{\bm{\chi}_{FI}}D] + 
        \overline{V^{\rm ext}_{F}} - \overline{V^{\rm ext}_{I}},
        \label{eq:avEfi}
\end{equation}
where $\overline{\bm{\chi}_{FI}}$ and $\overline{V^{\rm ext}_{X}}$ are

\[
\overline{\chi_{FI}}^{rs} = \frac{1}{2j_{F}+1}\sum\limits_{x\in F} F_{xx}^{rs} - 
\frac{1}{2j_{I}+1}\sum\limits_{x\in I} F_{xx}^{rs},
\]
\[
V^{\rm ext}_{X} = \frac{1}{2j_{X}+1}\sum\limits_{x\in X}V^{\rm
ext}_{xx},\ X {=} I ,\ F .
\]
Traces of the matrices $\left\langle f |h^{\rm eff}| f' \right\rangle$ and $\left\langle i |h^{\rm eff}|  i' \right \rangle$ are independent of the basis sets used. Let us compute them in the basis of functions with fixed values of the magnetic quantum number. We obtain
\footnote{
With the help of the Wigner-Eckart theorem (see~Ref.~\cite{Varshalovich:88}), one can write 
{$V^{\rm{ext}}_{xx}$, $x \in I,\ F$} as
\mbox{
\unexpanded{
$
        V^{\rm{ext}}_{X} = \sum\limits_{k} \langle x || V^{\rm{ext}}_k|| x \rangle
        \left ( \begin{array}{ccc} j_x& k& j_x \\ -m_x& 0& m_x \end{array}\right),
$
}}
where the reduced matrix element
\unexpanded{$ \langle x || V^{\rm{ext}}_k|| x \rangle$}
is independent of {$m_x$}.
It follows from the identity \unexpanded{$\sum\limits_m(-1)^{j-m}\left ( \begin{array}{ccc} j& k& j \\ -m& 0& m \end{array}\right)=\delta_{k0}\sqrt{2j+1}$}
that after averaging this expression over all the projection values {$m_x$} only the term with {$k=0$} survives.
}
\begin{equation}
        \overline{V^{\rm{ext}}_{X}} = U_{00}\, .
        \label{sec2:eq16}
\end{equation}
The spherical part $U_{00}$, depends only on the density of valence electrons in the region
$r>R_c$, and is the same for shells $I$ and $F$. Thus, the last two terms in expression~(\ref{eq:avEfi}) vanish. Matrix elements of the $\overline{\bm{\chi}_{FI}}$ operator are (see \appx{app_c})

\begin{equation}
        \begin{split}
                \overline{\chi_{FI}}^{rs} =
            \delta_{j_r,j_s}\delta_{m_r,m_s}(\overline{J}_{rs}({F})-\overline{J}_{rs}({I}) -\\
            -\overline{K}_{rs}({F})+\overline{K}_{rs}({I})),\\
            J_{rs}({X}) = \langle rx ||V_{0}||sx\rangle\sqrt{\frac{2j_r+1}{2j_x+1}},\\
            K_{rs}({X}) = \sum\limits_{k}\langle rx ||V_{k}||xs\rangle,\mbox{
при } {X}=F, {I}.
\end{split}
        \label{eq:chmatelem}
\end{equation}

The final expression for the transition energy in the relativistic average configuration approximation is

\begin{equation}
        \overline{\Delta E_{FI}} = E^{\rm Ic}_{F} - E^{\rm Ic}_{I} +
        \Tr[\overline{\bm{\chi}_{FI}}D].
        \label{sec2:eq17}
\end{equation}

Terms $E^{\rm Ic}_{F}$ and $E^{\rm Ic}_{I}$ do not depend by definition on the environment of the given atom and, therefore, they do not contribute to the chemical shift value. As a result, the chemical shift of the transition core energy in a compound $M$ with respect to that in the reference neutral atom $A$,  $\chi_{FI}(M,A)$, can be written as

\begin{equation}
	\begin{split}
        \chi_{FI}(M,A) = \overline{\Delta E_{FI}}(M) - \overline{\Delta E}_{FI}(A) =\\
 =  \Tr[\overline{\bm{\chi}_{FI}}\left[D(M){-}D(A)\right]].
	\end{split}
        \label{sec2:eq18}
\end{equation}

Thus, the chemical shift depends only on the change of the part of density matrix localized
 (on both variables)
in the region $r{<}R_c$, where the core transition takes place, and does not depend {\it directly} on any changes of the electronic densities out of this sphere.


%% file: sec3.tex
 \section{Calculation of chemical shifts in isolated atoms and ions}
   \label{sec6}
   \label{sec3n}
   \label{S_test}
The chemical shifts of $K{\alpha_1}$ and  $K{\alpha_2}$ transition energies in cations of group-14 metals Pb, Sn, Ge, and Si are studied with respect to the reference transition energies taken from calculations of the relativistic average ground-state configurations of given neutral atoms. Comparison of chemical shifts in calculations with frozen and relaxed Ic shells allows us (1) to estimate the computational errors, which arise as a result of neglecting the relaxation of Ic states of a given atom in its various compounds, and (2) to optimally divide the core electrons into Ic and Oc groups. The evaluation of chemical shifts is performed in few steps: 

\begin{itemize}
 \item  
Computation of transition energies using the Koopmans theorem.

 \item 
Partitioning the electrons into groups (see \sect{sec1}).

 \item  
Calculation of ionic states obtained by removing the electrons from the valence shells. In this calculation the inner core states are treated as frozen
  after
the reference state computation (and marked as ``frozen'' below in the text and tables).

 \item  
Computation of the cations with relaxed inner core shells.

 \item 
Evaluation of a chemical shift as the difference of the corresponding transition energies.
\end{itemize}

\input{tabl1}
In our calculations, the Ic groups were chosen to include all the states belonging to $1s-4f$ shells of Pb, $1s-3d$ shells of Sn, and $1s-2p$ shells of Ge and Si. The chemical shift values for the doubly and quadruply charged cations are listed in \Table{tabl1}. One can see from these data that the inner-core relaxation contributes only several percent of the absolute chemical shift values for the period four and heavier elements. We have also calculated the chemical shifts of $L$-transition energies in Pb$^{2+}$ compared with those in Pb (see \Table{tabl2}). The $L$-line energies are of much smaller magnitude than the $K{\alpha_1}$ and $K\alpha_2$ energies. Moreover, they are greatly influenced by relaxation of some core shells, which have the principal quantum numbers $n = n_v-2$, where $n_v$ stands for principal quantum number of the outermost valence shells. Such shells are usually considered as the Ic shells and can be frozen when studying chemical and spectroscopic properties.  However, in accordance with the partitioning rules given in \sect{sec1} these shells should be assigned to the Oc group at least for evaluating the given chemical shifts.

\input{tabl2}
The results presented in \Table{tabl2} demonstrate that one should account for relaxation of the $4d$ and $4f$ shells of Pb together with relaxation of the outer-more shells in order to evaluate the chemical shifts for $L$ XES lines in Pb with an accuracy within a few percents. Calculations of neutral Pb and Pb$^{2+}$ are performed in two ways: First, using the all-electron four-component atomic code {\sc hfd} \cite{HFD,Bratzev:77} that utilizes spherical symmetry and, second, employing the molecular spin-orbit direct configuration interaction code {\sc sodci} \cite{SODCI} together with the relativistic pseudopotential and core-restoration codes {\sc molgep} and {\sc core}. The {\sc molgep-core} codes implement the two-step approach \cite{Mosyagin:10a, Titov:06amin}) to study the properties of heavy-atom compounds, described by operators heavily concentrated in atomic cores, with moderate efforts. Comparing the results of different chemical-shift computations, one can estimate errors of the 
pseudopotential and core-restoration method advanced in the paper to include the chemical shift evaluation. 
Explicit treatment of the outer-core electrons in chemical-shift evaluation by the molecular code corresponds to a relaxed outer core in the case of using the atomic code. The results with and without explicit treatment of the outer core electrons are given in \Table{tabl3}. The computation of the configuration [Hg]$6p_{1/2}^2$ for the neutral Pb ground-state is used (with atomic code), whereas, the ground state configuration of Hg is used for the Pb$^{2+}$ case. The $jj$-coupling scheme is exploited in the Dirac--Fock calculations. The outer core $5s^25p^6 5d^{10}$ is treated as frozen after the calculation with nonrelativistic average configuration $6s^2 6p^2$ for the valence electrons.
\input{tabl3}

For a given state $\Psi$ of an isolated atom or its ion one can evaluate a mean value of the spin-angular projector $P_{lj} = \sum\limits_{m}|ljm\rangle\langle ljm|$, $N_{lj}[\Psi]=\langle\Psi|P_{lj}|\Psi\rangle$, that can be interpreted as the average number of electrons occupying all the one-electron states with some fixed quantum numbers $l$ and $j$.

\input{tabl4}

The $N_{lj}$ values are not necessarily integers in atomic correlation calculations, particularly, employing the $\Lambda{-}S$ coupling scheme. Having been obtained in {\sc sodci} calculations for Pb and Pb$^{2+}$, the $N_{lj}$ values are given in \Table{tabl4}. The $N_{lj}$ values obtained in correlation calculations of the ground state of a neutral Pb atom differ by less than 10\% by order of magnitude from integer numbers (corresponding to the one-configuration $jj-$coupling case) both for the frozen-outer-core treatment and for relaxed and correlated outer-core cases. The chemical shifts for the isolated Pb$^{2+}$ cation with respect to the neutral atom are approximately proportional to the difference between the corresponding values of $N_{p_{1/2}}$. Thus, we can conclude that the relative difference between chemical shifts obtained in various calculations is approximately equal to the difference between their $N_{p_{1/2}}$ values:
\[
\frac{\delta \chi_{FI}}{\chi_{FI}} \approx \left|\frac{\delta N_{p_{1/2}}}{N_{p_{1/2}}}\right|.
\]
The $\chi_{FI}$ quantity is a chemical shift obtained in the correlation calculation, the ${\delta \chi_{FI}}$ quantity is the difference between chemical shifts obtained in correlation and one-configuration calculations, and the $\delta N_{p_{1/2}}$ quantity is the difference between corresponding values of $N_{p_{1/2}}$. 

\input{tabl5}
It is clear from the results for chemical shifts listed in \Table{tabl3} that the difference between their values obtained in $jj$ and $\Lambda{-}S$ couplings is much greater than it could be expected from the $N_{lj}$ values given in \Table{tabl4} in the relaxed core case while it is consistent with the case of the frozen-outer-core calculation. This is due to interaction of the TC electrons with the outer-core electrons; the interaction gives a much greater contribution --- more than one order of magnitude in our case --- to the transition energy as compared with the energy of interaction of the TC electrons with valence electrons. In the case of the one-configuration atomic calculation, one can represent the density matrix as a direct sum of the valence $\rho_{\rm{V}}$ and outer-core $\rho_{\rm{OC}}$ terms, $D=\rho_{\rm{V}} + \rho_{\rm{OC}}$. In particular, the contributions from these terms to the $K\alpha_1$ and $K\alpha_2$ transition energies in the neutral and double-charged lead atom are listed in \Table{tabl5}.
One can see from these results that the contribution from the interaction of TC electrons with outer-core electrons is almost independent of the atomic partial charge as opposed to that with the valence electrons. The chemical-shift value (computed as a difference of the mean values of a given one-electron operator in the cases of the neutral atom and its cation) is the residual between two very close values, each of them is determined with an error of 10\% by the order of magnitude in calculations given in \Table{tabl5}. Taking into account the outer core contributions to the XES chemical shifts on a high and {\it identical} level of accuracy (quality) in calculations of different heavy-atom systems is a challenging problem for modern relativistic quantum chemistry because of limited sizes of the used basis sets (leading to the basis-set-superposition error problems, see Ref.~\cite{Duijneveldt:94} and references therein) and limited levels of correlation treatment in practical calculations of polyatomic systems which are of mainstream interest. The discussed outer-core contributions are rather large and difficult to control by absolute values in different types of chemical bonding of a given atom with others.
Such kinds of calculation of chemical shifts between molecules or/and periodic structures may be numerically unstable due to this fact
because of the inevitable use of the approximate computational methods for many-electron systems. To minimize the uncertainty, the contribution from the outer-core relaxation must be taken into account by some way {\it directly} (e.g., by using the perturbation technique for outer core, etc.) and not as a small difference between two big values.


%% file: tabl1.tex
\begin{table}[h!]
    \caption{
Chemical shifts of $K{\alpha_1}$ and $K{\alpha_2}$ lines for the doubly charged cations of group-14 elements, Pb$^{2+}$, Sn$^{2+}$, Ge$^{2+}$, Si$^{2+}$, and corresponding quadruply charged cations with respect to the neutral atoms.$^c$}
\scalebox{0.8}{%
\begin{tabular}{lcccccc}
\toprule
        &\multicolumn{6}{c}{\mbox{A$^{2+}$} with respect to A.}\\
        \hline
           &$\chi_{K_{\alpha1}}$, meV\footnotemark[1] &$\chi^{\rm{fr}}_{K_{\alpha1}}$, meV\footnotemark[2]&$\delta_{K_{\alpha1}}$,\%&$\chi_{K_{\alpha2}},$ meV\footnotemark[1]&$\chi^{\rm fr}_{K_{\alpha2}}$, meV\footnotemark[2]& $\delta_{K_{\alpha2}}$, \%\\
        \hline
        Pb & 127  & 130&2.3  & 150 & 150 & 0.3\\
        Sn & 116  & 123&5.1  & 170 & 166 & 2.3\\
        Ge & 214  & 228&6.5  & 228 & 214 & 6.1\\
        Si & 671  & 981&46   & 981 & 671 & 31 \\
        \hline
        &\multicolumn{6}{c}{\mbox{A$^{4+}$} with respect to  A.}\\
        \hline
           &$\chi_{K_{\alpha1}}$, meV\footnotemark[1] &$\chi^{\rm{fr}}_{K_{\alpha1}}$, meV\footnotemark[2]&$\delta_{K_{\alpha1}}$,\%&$\chi_{K_{\alpha2}},$ meV\footnotemark[1]&$\chi^{\rm fr}_{K_{\alpha2}}$, meV\footnotemark[2]& $\delta_{K_{\alpha2}}$, \%\\
        \hline
        Pb & 347 & 359 & 3.4 & 362 & 355 & 1.9 \\
        Sn & 379 & 400 & 5.5 & 441 & 423 & 4.1 \\
        Ge & 741 & 789 & 6.4 & 741 & 789 & 6.4 \\
        Si &1990 &3538& 77.2 &3538 &1990 & 77.2\\
\toprule
\end{tabular}
}
\footnotetext[1]{The results of all-electron calculations of the chemical shifts.}
\footnotetext[2]{The results of the chemical shift calculations with the frozen inner core.}
\footnotetext[3]{The relative errors arising from neglecting the inner core relaxation are $\delta_I =\left|\,\frac{\chi_{I}^{\rm{fr}}-\chi_{I} }{\chi_{I}}\,\right|$\,. In the frozen-core calculations, the inner cores of Pb and Sn cations were taken from evaluation of the relativistic average configuration for the Pb and Sn atoms, and from the non-relativistic average configurations for the Ge and Si atoms. In these studies, the inner core shells are  $1s{-}4f$ for Pb, $1s{-}3d$ for Sn, $1s{-}2p$ for Ge  and $1s{-}2p$ for Si.}
\label{tabl1}
\end{table}

%% file: tabl2.tex
\begin{table}[!h]
\caption{The chemical shift values of XES $L$ lines of Pb$^{2+}$ with respect to the neutral Pb atom. The N$^{\mathrm{fr}}{=}0$ column corresponds to computations with relaxation of all electronic shells; [Kr], [Kr]$4d^{10}$ and [Kr]$4d^{10}4f^{14}$ denote computations with the frozen shells from $1s$ to $4p$, $4d$ and $4f$, correspondingly.}
\begin{ruledtabular}
\begin{tabular}{lccccc}
Frozen shells: & N$^{\rm fr}{=}0$ & [Kr]& [Kr]$4d^{10}$ & [Kr]$4d^{10}4f^{14}$ \\
 \hline
 $\chi_{L_{\beta1}}$,  meV   &10 & 11& 18 &   21 \\ 
 $\chi_{L_{\alpha2}}$, meV   &30 & 30& 37 &   39 \\
 $\chi_{L_{\alpha1}}$, meV   &35 & 34& 42 &   44 \\
\end{tabular}
\end{ruledtabular}
\label{tabl2}
\end{table}

%% file: tabl3.tex

\begin{table}[h!]

	\caption{The chemical-shift values of the Pb$^{2+}$ $K{\alpha_1}$ and $K{\alpha_2}$ lines with respect to the neutral Pb atom.}
	\label{tabl3}
	\begin{ruledtabular}
	\begin{tabular}{lcccc}
	line	 & $\chi^{\rm{A}}$, meV\footnotemark[1]&$\chi^{\rm{M}}$, meV\footnotemark[2]&$\chi^{\rm{A,\,fr}}$, meV\footnotemark[3]& $\chi^{\rm{M,\, fr}}$, meV\footnotemark[4]\\
        \hline
        $K_{\alpha1}$ &{} 130          &{} 88       &{} 145        &{} 145   \\ 
        $K_{\alpha2}$ &{} 150          &{} 33       &{} 165        &{} 156   \\
	\end{tabular}
	\end{ruledtabular}
	\footnotetext[1]{Chemical shift calculated with the atomic code {\sc hfd} \cite{HFD}; the outer core is unfrozen.}
	\footnotetext[2]{Chemical shift calculated with the atomic code {\sc hfd} \cite{HFD}; the outer core is frozen.} 
	\footnotetext[3]{Chemical shift calculated with the molecular code {\sc molgep/sodci} \cite{MOLCAS,SODCI,Titov:01};  the outer core is unfrozen.}
	\footnotetext[4]{Chemical shift calculated with the molecular code {\sc molgep/sodci} \cite{MOLCAS,SODCI,Titov:01}; the outer core is frozen.}
\end{table}

%% file: tabl4.tex

\begin{table}[h!]
        \centering
		\caption{The occupation numbers $N_{lj}$ of one-electron $p_{1/2}$ and $p_{3/2}$ states of the neutral Pb atom and Pb$^{2+}$ cation in the ground states evaluated as the mean values of the spin-angular projectors $P_{lj} = \sum\limits_{m}|ljm\rangle\langle ljm|$ for 22-electron {\sc molgep/sodci} calculations with frozen $5p$ from ionic Pb ground state (``OC-frozen'') and relaxed $5p$ (``OC-relaxed'').}
		\begin{ruledtabular}
                \begin{tabular}{lcccc}
			&\multicolumn{2}{c}{Pb}&\multicolumn{2}{c}{Pb$^{2+}$}\\
                       & $N_{p_{1/2}}$ &$N_{p_{3/2}}$ & $N_{p_{1/2}}$ &$N_{p_{3/2}}$ \\
                    \hline
                  OC-frozen\footnotemark[1]
                            &  $1.8$      &$0.2$ & $0$ & $0$ \\
                  OC-relaxed&  $3.9$ &$4.1$    &$2.0$ & $4.0$  	\\
                \end{tabular}
		\end{ruledtabular}
        \label{tabl4}
\footnotetext[1]{The occupation numbers of $5p_{1/2}$ and $5p_{3/2}$, 2 and 4, correspondingly, are added for comparison with the OC-relaxed case.}

\end{table}

%% file: tabl5.tex
\begin{table}[h!]
		\caption{Contributions to the energies of $K{\alpha_1}$ and $K{\alpha_2}$ transitions arising from interaction of the TC electron with the valence ($6s, 6p$) and outercore ($5s, 5p, 5d$) electrons.}
		\begin{ruledtabular}	
                \begin{tabular}{llcc}
                               &                                                   & Pb     &Pb$^{2+}$\\
                        \hline
                        $K_{\alpha1}$ & $\varepsilon^{\rm OC}$, eV & 13.788 & 13.816\\
                                                       & $\varepsilon^{\rm V}$, eV  &0.543   &  0.387\\
                        \hline
                        $K_{\alpha2}$ & $\varepsilon^{\rm OC}$, eV &11.538  &11.561\\
                                      & $\varepsilon^{\rm V}$, eV  &0.501   & 0.327 \\
                \end{tabular}
		\end{ruledtabular}
        \label{tabl5}
\end{table}

%% file: sec4.tex
\section{Chemical shifts of ${\rm Pb}$ $K{\alpha_1}$ line in {PbO} and {PbF}$_{2}$ molecules}
\label{sec4n}

\input{tabl6}
We have studied the chemical shifts of $K{\alpha_1}$ transition energies on Pb in the PbO and PbF$_2$ molecules with respect to the neutral lead atom (the experimental value of neutral Pb $K{\alpha_1}$ line energy is $74.97011$ keV~\cite{Deslattes:03}). The oxidation number of Pb in these molecules is $+2$, therefore, their chemical shifts are compared to those of Pb$^{2+}$ and presented in \Table{tabl6}. The results of calculations differ strongly for the molecules compared to Pb$^{2+}$ since the chemical bonds in both molecules are not purely ionic but have notable covalent admixtures, particularly in the oxide. Partial Mulliken and electronegativity-based charges of lead in PbO are $+0.9$ and $+0.86$ \cite{bacanov:62}, respectively. However, the change of the total charge of $p$ electrons is much less in the area significant for the $K\alpha_1$ line chemical shift values. The considered area is limited by the sphere with radius $R_c{=}0.5$ a.u.\ and $R_c$ is selected in such a way that the $2p$ shell of Pb has negligible density outside the sphere.

To analyze particular contributions to the chemical shift in compounds of Pb compared with atomic lead given in \Table{tabl6}, we introduce the quantities $q^{<}_{lj}$ which describe the partial wave charges (corresponding to the total electronic densities for the states with fixed $l$ and $j$) concentrated in the spherical region with radius $R_c{=}0.5$~a.u., in which the one-center restoration of electronic structure in the Pb core is performed. These values are calculated by the following expression for the compound M:
\begin{equation}
q^<_{lj}(M) = Tr[D(M)P_{lj}],
\end{equation} 
where $D(M)$ is the valence and outer core electrons density matrix restricted to the spherical region (see \appx{app_a}), and $P_{lj}$ is the spin-angular projector $P_{lj} = \sum\limits_{m=-j}^{j} \left|ljm\right\rangle\left\langle ljm \right |$ on the states with given quantum numbers $l$ and $j$. 

One can see from \Table{tabl6} that significant contribution to the chemical shift, at the level of 10\% by order of magnitude, is coming from interaction of TC electrons with those occupying the perturbed $6s$ states. The fraction of the latter in the atomic core is increased due to disappearance of the electronegative potential from $6p$ electrons in this region in the case of Pb$^{2+}$, while the perturbation of $6s$ states is relatively small in the PbO case due to a high covalent share in the Pb--O bond and influence of the valence electrons of oxygen.

To estimate the chemical-shift contribution from the atomic orbitals of oxygen, we have also presented the results of evaluation of chemical shifts without atomic oxygen orbitals taken into account at the core-restoration stage. The molecular calculations are performed within the GRECP-configuration-interaction method \cite{Mosyagin:10a} by using the {\sc molgep-sodci}
 codes \cite{MOLCAS, SODCI, Titov:01}.

The study is performed with the frozen outer core of Pb including the $5s^25p_{1/2}^25p_{3/2}^4 5d_{3/2}^4 5d_{5/2}^6$ shells. The interatomic distance is $4.2$~a.u.\ according to the experimental datum for the PbO molecule given in Ref.~\cite{CRC:10}. The distance between Pb and F atoms in PbF$_2$ is taken to be $4.2$~a.u.\ \cite{CRC:10}; however, we have considered the linear geometry of PbF$_2$ here assuming the comparison with further chemical shift measurements in cubic crystalline PbF$_2$.

The experimental chemical shift of the Pb $K{\alpha_1}$ line in the crystalline PbO with respect to that of metallic lead is $54\pm8$ meV \cite{Egorov:92}. It should be emphasized that one can compare atomic and molecular Pb and PbO computations with the experimental solid-state data only qualitatively, since the difference in the electronic structures of the atomic and metallic lead as well as of crystalline and molecular PbO is significant. Nevertheless, we can conclude that the agreement between experimental data and our results is satisfactory; both the experimental and theoretical results differ notably from the chemical-shift values based on the ionic model and Mulliken occupancy analysis that gives the Pb partial charges $\sim 0.9$ in PbO and $\sim 1.5$ for PbF$_2$.

%% file: tabl6.tex
\begin{table}[!h]
         \caption{ The $K{\alpha_1}$ chemical shifts and partial wave charges values in lead compounds.$^a$}
\begin{ruledtabular}
    \begin{tabular}{lcccc}
           &   $100q_s^<$ & $100q_{p_{1/2}}^<$ & $100q_{p_{3/2}}^<$& $\chi_{K_{\alpha1}}$, meV\\
          \hline
Pb         &  1.814             &   0.903                 &    0.1     &     \\       
Pb$^{2+}$  &  2.162             &   0                     &     0      & $145$  \\
\hline
PbO\footnotemark[1]       &  1.767             &   0.432                 &    0.456   & $40$  \\
PbO\footnotemark[2]  &  1.789             &   0.381                 &    0.496   & $41$  \\
PbO\footnotemark[3] &  1.789             &   0.455                 &    0.503   & $23$  \\
PbO (exp. value)\footnotemark[4] &                    &                         &            & $54\pm 8$\\
\hline
PbF$_{2}$        &  1.950             &   0.153                 &    0.254   & $85$\\
    \end{tabular}
\end{ruledtabular}
 \label{tabl6}
\footnotetext[1]{$q^{<}_{lj}$ is the part of total charge of all the valence electron shells with given orbital and total-angular-momentum (``partial waves'') numbers, within a sphere of radius $ R_c = 0.5 $~a.u.\ and centered on the lead nucleus;
$\chi_{K{\alpha_1}}$ is the chemical shift of $K{\alpha_1}$ line with respect to the neutral Pb atom.}
\footnotetext[2]{Calculations with the oxygen basis set taken from Ref.~\cite{Petrov:05a}.}
\footnotetext[3]{Calculations with the extended oxygen basis set.}
\footnotetext[4]{Calculations with neglecting the oxygen orbital contributions at the core-restoration stage.}
\footnotetext[5]{The experimental chemical-shift value of the $K{\alpha_1}$ lead XES lines in the PbO crystal with respect to the crystalline metallic lead (see the discussion in text).}
\end{table}

%% file: conc.tex

\section*{Conclusions}

\sloppy

A method of evaluating the XES-line chemical shift is developed. This method can be used to study the electronic transitions in cores of elements starting from period four of the Periodic Table and below. An analytic expression for the chemical shift as a difference of mean values of a proposed effective one-electron operator being calculated in two systems containing a given element is obtained. It is shown that the influence of changes in the electronic densities outside the atomic core, where the TC electrons are localized, on the XES chemical shift values  is mainly negligible.

The expression for the chemical shift value is obtained in the relativistic configuration average approximation, the sudden-transition and frozen-inner-core approximations are also employed. Applying the sudden-transition model, one neglects the relaxation of the valence and outer-core electron densities during the transition. This approximation is good enough for the issues considered, because typical transition times for inner-core electron are $10^{-16}{\div}10^{-13}$~s, whereas the transition (relaxation) times for the outer-core and valence electrons are $\tau' \sim 10^{-12}{\div}10^{-8}$~s~\cite{Labzowsky:96}. Using the frozen-core approximation for the inner-core one-electron states of a given atom we assume that their changing in the atom is negligible when its chemical environment is varies from one compound to the other.
The relaxation effects due to the x-ray-induced ejection of a core electron preceding the considered core transition is also neglected. Note that using the one-electron approximation for the inner-core electrons does not assume the same level of treatment for outer core and valence electrons which can be explicitly correlated. The latter is particularly important to
   optimize the computational efforts for chemical-shift evaluation in
the systems with complicated valence structure in contrast to other known theoretical approaches (mainly based on DFT).

Corrections to the proposed chemical-shift expression restricted by the relativistic average configuration approximation are analyzed. They are shown to be mostly a few percents compared to the average chemical shift values.

Atomic calculations of chemical shifts of $K{\alpha_1}$ and $K{\alpha_2}$ lines for the group-14 transition metal cations compared to the neutral atoms are performed with the {\sc hfd} code \cite{HFD, Bratzev:77}. There are two variations used in our calculations, either with or without taking account of the inner-core relaxation. As one can see from the results, neglecting the inner-core relaxation leads to a few percent increase of the chemical-shift errors for group-14 elements starting from Ge. However, the chemical-shift error for Si is 78\% because the partitioning of the core shells onto the inner- and outer-core subspaces is meaningless here.

Study of $L$ XES chemical shifts on the lead cations with respect to the neutral Pb atom is performed. It is shown that one needs to account for relaxation of the $4d$ and $4f$ core shells to attain the level of accuracy for chemical-shift values within 10\%.

The GRECP-configuration-interaction study of the $K{\alpha_1}$ line chemical shifts for Pb$^{2+}$ with respect to the neutral Pb atom as reference is performed with the {\sc sodci} code. The errors due to the electronic structure restoration approximation used for the valence and outer core orbitals in atomic cores are estimated.

The GRECP-configuration-interaction calculations of the $K{\alpha_1}$ line chemical shifts on Pb in PbO and PbF$_2$ with respect to the neutral atom are also performed. The obtained chemical shifts for PbO are in a reasonable agreement with available experimental data for the crystallic PbO.
%

\section*{Acknowledgments}
We are grateful to V.V.~Fedorov for the initiation of this work and useful comments, to A.V.~Zaitzevskii, N.S.~Mosyagin, A.N.~Petrov, L.V.~Skripnikov, A.D.~Kudashov for many important remarks, and to A.A.~Petrunin, A.E.~Sovestnov, and E.V.~Fomin for discussions. 
   This research was partially supported by the RFBR Grant~No.~13-03-01307a

%% file: app_a.tex

\section{Partitioning the one-electron density submatrices into the ``outer'' and ``inner'' parts. The external electric field in the atomic core}
\label{app_a}

Let us define the one-electron  W$-$group density submatrices $D,\ \bm{\rho}_\mathrm{W}^{<>},\ \bm{\rho}_\mathrm{W}^{><}$ and $\bm{\rho}_\mathrm{W}^{>>}$ in coordinate representation using step functions $\theta_<(\vec{r}){=}\theta(R_c{-}|\vec{r}|)$ and $\theta_>(\vec{r}){=}1{-}\theta_<(\vec{r})$:
\begin{equation}
\begin{split}
        \bm{\rho}_\mathrm{W}(\vec{r},\vec{r'}) = D (\vec{r},\vec{r'}){+}\bm{\rho}_\mathrm{W}^{<>}(\vec{r},\vec{r'}){+}\bm{\rho}_\mathrm{W}^{><}(\vec{r},\vec{r'}){+}\bm{\rho}_\mathrm{W}^{>>}(\vec{r},\vec{r'})\,,\\
    D(\vec{r},\vec{r'})\equiv \bm{\rho}_\mathrm{W}^{<<}(\vec{r},\vec{r'}) =
     \theta_<(\vec{r}) \bm{\rho}_\mathrm{W}(\vec{r},\vec{r'}) \theta_<(\vec{r'})\, ,\\
    \bm{\rho}_\mathrm{W}^{<>}(\vec{r},\vec{r'}) = \theta_<(\vec{r}) \bm{\rho}_\mathrm{W}(\vec{r},\vec{r'})\theta_>(\vec{r'})\, ,\\
    \bm{\rho}_\mathrm{W}^{><} (\vec{r},\vec{r'})= \theta_>(\vec{r}) \bm{\rho}_\mathrm{W}(\vec{r},\vec{r'})\theta_<(\vec{r'})\, ,\\
    \bm{\rho}_\mathrm{W}^{>>} (\vec{r},\vec{r'})= \theta_>(\vec{r})\bm{\rho}_\mathrm{W}(\vec{r},\vec{r'})\theta_>(\vec{r'})\, .
 \end{split}
 \label{app_v:eq01}
\end{equation}
Consider the off-diagonal submatrices $\bm{\rho}_{\mathrm{WR}} =P_\mathrm{W} \bm{\rho} P_\mathrm{R}$ and $\bm{\rho}_{\mathrm{RW}} =P_\mathrm{R} \bm{\rho} P_\mathrm{W}$ defined in \Eq{sec2:rhodef} in coordinate representation. 
For $\bm{\rho}_{\mathrm{WR}}(\vec{r},\vec{r'})$, the radius vector $\vec{r}$ corresponds to states of group W and $\vec{r'}$
to states of group R; for $\bm{\rho}_{\mathrm{RW}}(\vec{r},\vec{r'})$, $\vec{r}$\  corresponds to R, and $\vec{r'}$
to W.  We define the submatrices $\bm{\rho}_\mathrm{WR}^<$ and $\bm{\rho}_\mathrm{WR}^>$ such that $\bm{\rho}_\mathrm{WR}= \bm{\rho}_\mathrm{WR}^>+ \bm{\rho}_\mathrm{WR}^<$ analogously to \Eq{app_v:eq01}:
\[
\begin{split}
        \bm{\rho}_\mathrm{WR}^{<}(\vec{r},\vec{r'}) \equiv \bm{\rho}_\mathrm{WR}^{<>} = \theta_<(\vec{r})\bm\rho_\mathrm{WR}(\vec{r},\vec{r'})\, ,\\
        \bm{\rho}_\mathrm{WR}^{>}(\vec{r},\vec{r'}) \equiv \bm{\rho}_\mathrm{WR}^{>>} =  \theta_>(\vec{r})\bm\rho_\mathrm{WR}(\vec{r},\vec{r'})\, . 
\end{split}
\]
Similarly we also define $\bm{\rho}_\mathrm{RW}^{~~<}$ and $\bm{\rho}_\mathrm{RW}^{~~>}$.

Furthermore, substitute $\bm{\rho}_\mathrm{WR}{=} \bm{\rho}_\mathrm{WR}^{<}{+}\bm{\rho}_\mathrm{WR}^{>}$ and $\bm{\rho}_\mathrm{RW}{=} \bm{\rho}_\mathrm{RW}^{~~<}{+}\bm{\rho}_\mathrm{RW}^{~~>}$ into \Eqs{sec2:eq10} and (\ref{sec2:eq8})
and consider the traces of product of operators $\bm{F}_{xx}$ with these submatrices in coordinate representation.
Taking into account that the one-electron states of group R have negligible densities in the Ic region of atom $A$ (see Sec.~\ref{sec2}), we can write
\begin{equation}
        \Tr[\bm{F}_{xx}\bm{\rho}^<_{WR}] = \int\limits_{\vphantom{\vec{'}}|\vec{r}|<R_c}d\vec{r}\,\int\limits_{|\vec{r'}|>R_c}d\vec{r'}F_{xx}(\vec{r'},\vec{r})\bm{\rho}_{WR}(\vec{r},\vec{r'}),
\label{eq:coord_spur}
\end{equation}
where $\bm{F}_{xx}(\vec{r},\vec{r'})$ is defined as a sum of Coulomb and exchange terms, correspondingly: 
\[
\begin{split}
\bm{F}_{xx}(\vec{r},\vec{r'}) = 
 \delta(\vec{r}{-}\vec{r'}) \int d\vec{r_1}\, [V(\vec{r_1}{-}\vec{r'})|\varphi_x(\vec{r_1})|^2]\,\\\
-\ \varphi^*_x(\vec{r})\varphi_x(\vec{r'})V(\vec{r}{-}\vec{r'})\ .
\end{split}
\]
 One can easily realize that expression~(\ref{eq:coord_spur})
  vanishes
because the integration domains over $\vec{r}$ and $\vec{r'}$ are not overlapping
and the TC state $\varphi_x(\vec{r})$
  amplitude is vanishing
outside the sphere with radius $R_c$. Analogously, the expressions for chemical shift with the submatrices $\bm{\rho}^{~~<}_{RW}$, $\bm{\rho}^{<>}_{W}$ and $\bm{\rho}^{><}_{W}$ will also be zero.

Thus, only the following parts of the total density matrix can contribute to the chemical shift value: $\bm{\rho}_\mathrm{R}$,
 $\bm{\rho}_\mathrm{WR}^{>}$, $ \bm{\rho}_\mathrm{RW}^{~~>}$
$\bm{\rho}_\mathrm{W}^{>>}$, $\bm{\rho}_\mathrm{Ic}$ and $D$.

Let us represent the contribution to TC energy arising from the interaction of the TC electron with nuclei of other atoms and the electronic components corresponding to $\bm{\rho}_\mathrm{R}$, $\bm{\rho}_\mathrm{WR}^{>}$, $\bm{\rho}_\mathrm{RW}^{~~>}$ and $\bm{\rho}_\mathrm{W}^{>>}$ submatrices
with the domains localized outside the sphere: $|\vec{r}|,|\vec{r'}|>R_c$
as action of a ``crystal'' (or ``molecular'') external field operator $V^{\rm{ext}}$:

\begin{equation}
	\begin{split}
        V^{\rm{ext}}_{xx} = \sum_{A'} V^{\rm{A'}}_{xx} + \\
        + \sum\limits_{rs}(\bm{\rho}_\mathrm{R}{+}\bm{\rho}_\mathrm{WR}^{>}{+}\bm{\rho}_\mathrm{RW}^{~~>}
        +\bm{\rho}_\mathrm{W}^{>>})_{rs}(V_{xxrs}-V_{xrxs}),\\ \mbox{ $x=i,f$}.\ \,
	\end{split}
  \label{app_v:eq11}
\end{equation}
With appropriate choice of $R_c$, when the tails of the core states in the region with $|\vec{r}|>R_c$ become small enough, the exchange terms $V_{xrxs}$ are negligible and $V^{\rm{ext}}$ inside the sphere becomes a local operator that can be written as
a multipole expansion on the nucleus $A$:
\begin{equation}
        V^{\rm{ext}}(\vec{r}) \approx \sum_{km} U_{km}r^k Y_{km}(\Omega) \, ,\ \quad \mbox{  $r<R_c$\ .}
  \label{app_v:eq12}
\end{equation}

%% file: app_b.tex

%
\section{First-order perturbation theory analysis of the errors arising from assumption of equiprobability of the TC electron transitions.}

Evaluating a chemical shift for core-to-core transition energies, we average them over all the projections of total angular momenta both for the initial and final one-electron states. As is shown in \sect{sec2}, the chemical-shift contributions arising from interactions of TC electrons with electronic densities localized outside the sphere with radius $R_c$ (centered on the nucleus of a given atom $A$ or all the other atoms in the considered system) vanish in this case. Such an averaging corresponds to a situation in which all the transitions between $I$ and $F$ shells have equal probabilities. We can estimate the errors arising from this approximation when considering only the electric-dipole transitions for simplicity. Then, after averaging the initial and final states over all the total-angular-momentum projections, $m_i$ and $m_f$, we have to take into account the constraint on possible values of $m_i$ and $m_f$:

\begin{equation}
    |m_i-m_f|\le 1\ .
    \label{eq:av0}
\end{equation}
Note, however, that the electric-dipole approximation works well for light atoms. For compounds containing heavy atoms, the magnetic and higher multipole electric transitions become significant due to relativistic effects and perturbation of the spherical symmetry of atomic cores in compounds; therefore, the constraint \Eq{eq:av0} is weakened in practice. Thus, using \Eq{eq:av0} we can only estimate the upper bound on the errors arising from assumption of equal probabilities for the transitions between $I$ and $F$ shells with different $m_i$ and $m_f$.

Let us estimate the difference in the average transition energies evaluated with and without condition (\ref{eq:av0}). Consider the case, when all the final and initial states are eigenvectors of the projection of the total-angular-momentum operator, which can have different energies:

\begin{equation}
        \overline{\Delta E_{FI}} = \frac{1}{N}\sum\limits_{|m_i-m_f|\le 1}
        \varepsilon_{F;m_f}-\varepsilon_{I;m_i}\ ,
        \label{eq:av1}
\end{equation}
where $\varepsilon_{I;m_i}$ and $\varepsilon_{F;m_f}$ are energies of the one-electron states belonging to the $I$ and $F$ shells, with the projections $m_i$ and $m_f$, correspondingly. The  coefficient $N$ takes the following values

\begin{equation}
        \left\{ 
        \begin {array}{rcr}
        N &=& 3(2j_{I}-1)+ 4\mbox{, }j_{I}=j_{F},\\
        N &=& 3(2\min(j_{I},j_{F})+1)\mbox{,  } j_{I}\ne j_{F}
 \end{array}\right.
        \label{eq:av2}
\end{equation}
The constraint~(\ref{eq:av1}) does not influence on the averaging over $m_i$ and $m_f$ only for $J_I = J_F = \frac{1}{2}$, otherwise, the independent and constrained averages do not coincide. Let us write the energies of one-electron states $\varepsilon_i$, $\varepsilon_f$ as

\begin {equation}
\varepsilon_{x}=\overline{\varepsilon}_{X} + (\varepsilon_{x}
-\overline{\varepsilon}_{X}),
\end{equation}
where $\overline{\varepsilon}_X = \sum\limits_{x\in X}\varepsilon_x$ is the average orbital energy for shell $X= I,\ F$.

After substituting $\varepsilon_{X,m_x}$ into expression~(\ref{eq:av1}) we obtain

\begin{equation}
        \begin{split}
                \overline{\Delta E_{FI}} = (\overline{\varepsilon}_{F}-\overline{\varepsilon}_{I}) + \delta\varepsilon_{FI}\, ,\\ 
	\delta\varepsilon_{FI} = - \frac{1}{N}\sum\limits_{ |m_f|>j_I, \atop |m_f\pm
	1|>j_I}(\varepsilon_{F,m_f} - \overline{\varepsilon}_{F}).
\end{split}
        \label{eq:av3}
\end{equation}
Note that the dependence of $\delta\varepsilon_{FI}$ on the initial $I$ shell arises from dependence of the summation index limits in the above equation on the total angular momentum $j_I$.  

Consider the matrix elements of effective one-electron Hamiltonian \Eq{sec2:eq8mod} , $h^{\rm eff}_{xx'} \equiv \left\langle x | h^{\rm eff} | x' \right\rangle$, 
where $|x\rangle$ and $|x'\rangle$ are eigenvectors of the total angular momentum projection operator, they belong to the Ic shell $X$ with a fixed total angular momentum $j_{X}, X=I,F$. According to the Wigner-Eckart theorem, matrix elements of $h^{\rm eff}_{xx'}$ can be written in the form:

\begin{equation}
        h^{\rm eff}_{xx'} =
        \sum\limits_{0\le k\le 2 j_x}
        \left \langle x || h^{\rm eff}_k || x' \right\rangle\left (
        \begin{array}{ccc} j_X& k& j_X \\ -m_x& k& m_{x'} \end{array}\right)\ ,
        \label{eqpt1}
\end{equation} 
where $m_x$ and $m_{x'}$ are projections of total angular momentum of the corresponding states, 
$\left \langle x || h^{\rm eff}_k || x' \right\rangle = \left \langle x || V^{\rm ext}_k || x'\right\rangle{+}\left \langle x || \Tr[\bm{F_k} D] || x' \right\rangle$ are the conventional reduced matrix elements \cite{Varshalovich:88}
  for the $k$-rank operators given in the spherical tensor representation,
and the terms $\left \langle x || V^{\rm ext}_k || x'\right\rangle$ and $\left \langle x || \Tr[\bm{F_k} D] || x' \right\rangle$ are defined below.

\sloppy

Only the terms with $k\ne0$ contribute to the average orbital energy correction $\delta\varepsilon_{FI}$. Estimate the $\left \langle x || V^{\rm ext}_k || x' \right\rangle$ and $\left \langle x || \Tr[\bm{F_k} D] || x' \right\rangle$ terms by their order of magnitude.
The external field (\ref{app_v:eq12}) $V^{\rm ext} = \sum\limits_k U_{km}r^k Y_{km}(\Omega)$ is created by the other atoms of a compound and the electronic density outside the sphere with radius $R_c$ centered on the considered atom. For electroneutral compound, $U_{km}$ can be estimated as $\lesssim Q_V/R_V^{k+1}$, where $Q_V$ is the valence of the given atom and $R_V>R_c$ is the radius of its valence shells. 

Let us write matrix elements $\langle x|| V^{\rm ext}_k || x' \rangle$ as
\begin{equation}
\langle x || V^{\rm ext}_k || x' \rangle = U_{km} R^{<>}_{kX} \left ( \begin{array}{ccc} l_X& 2& l_X \\ 0& 0& 0 \end{array}\right),
\label{app_b:eq7}
\end{equation}
where the radial integrals $R^{<>}_{kX}$ are defined as\footnotemark
\[
R^{<>}_{kX} =\int\limits_{r<R_c}dr\,r^{k+2} |\varphi_{X}(r)|^2. 
\]
\footnotetext{%
One may estimate the upper bound of the radial integral 
$R^{<>}_{kX}$
 as
\[
R^{<>}_{kX}\le \langle r_X \rangle^k,
\] 
where $\langle r_{X} \rangle$ is the average radius of the $X$ shell. Thus,
\[
        \left \langle x || V^{\rm ext}_k || x' \right\rangle \lesssim \frac{Q_V}{R_V} \left(\frac{\langle r_X \rangle}{R_V}\right)^k. 
\]
}
  Consider the
contributions to $\delta\varepsilon_{FI}$ correction to the average energy of electronic transition from the shell with $j_I = 1/2$ to the shell with $j_F = 3/2$ of the lead atom (in standard notation: $p_{3/2}{\to}s_{1/2}$, $d_{3/2}{\to}p_{1/2}$) arising from
interaction with the external field. 
For $K$-series transitions ($np\to 1s$), the energy correction for the $s$ shell is zero, whereas the correction for the $p$ shell is determined by the matrix elements

\[
\left\langle x |V^{\rm ext}| x' \right\rangle =
R^{<>}_{k{=}2,X}\sum\limits_{m=-2}^2\left \langle x | Y_{km} | x' \right\rangle.
\]

Let us estimate now contributions to the $\delta\varepsilon_{FI}$ correction from the terms $\left \langle x || \Tr[\bm{F}_k D] || x' \right\rangle$ by their order of magnitude and, therefore, consider only the direct Coulomb interaction terms with assessment
$\left \langle x || \Tr[\bm{F}_k D || x' \right\rangle \sim \left \langle x || \Tr[\bm{J}_k D] || x' \right\rangle$.

Write $ \left \langle x || \Tr[\bm{J}_k D] || x' \right\rangle$ analogously to \Eq{app_b:eq7}:
\begin{equation}
        \left \langle x || \Tr[\bm{J}_k D] || x' \right\rangle \sim R^{<<}_{kX}\ ,
        \label{eq:coulsim}
\end{equation}
where the radial integral $R^{<<}_{kX}$ is

\begin{equation}
R^{<<}_{kX} = \int\limits_{r_1<R_c} dr_1\int\limits_{r_2<R_c} dr_2
\frac{r_>^{k+2}}{r_<^{k-1}}|\varphi_{X}(r_1)|^2\rho_V(r_2)\ ,
\label{eq:rll}
\end{equation}
and the valence electron density $\rho_V(r)$ is 

\[
        \rho_V(r) = \sum_{rs}D_{rs}\varphi_r(r)\varphi^*_s(r)\ .
        \label{eq:rhoV}
\]
\input{tabl7}

 The values  of the radial integrals $R^{<>}_{kX}$ and $R^{<<}_{kX}$ for the inner-core shells of Pb in the case of $k=2$ are given in \Table{tabl7}.

To derive the final estimation for correction to the average transition energy
one has to take into account the value of $N^{-1}$ in expression (\ref{eq:av1}):

\begin{equation}
        \begin{split}
        \delta\varepsilon_{FI}{\sim}
        \frac{2}{N}(\frac{Q_V}{R_V^3}R^{<<}_{2F}{+}R^{<>}_{2F})\left ( \begin{array}{ccc} j_F& 2& j_F \\ j_F& 0& -j_F \end{array}\right)\left( \begin{array}{ccc} j_V& 2& j_V \\ -j_V& 0& j_V \end{array}\right)\cdot\\
        \cdot\left ( \begin{array}{ccc} l_F& 2& l_F \\ 0& 0& 0 \end{array}\right)\left ( \begin{array}{ccc} l_V& 2& l_V \\ 0& 0& 0 \end{array}\right),
        \end{split}
        \label{eq:finav}
\end{equation}
where $j_F$ and $l_F$ are  the total and orbital angular momenta of the final TC shell, $j_V$ and $l_V$ are quantum numbers for the most populated valence shell. The corresponding product of the angular multipliers and $\frac{2}{N}$ is $\sim1/50$ by the order of magnitude for $p$ and $d$ shells. Multiplying the latter by the radial integrals $R^{<>}_{kX}$ and $R^{<<}_{kX}$, we obtain that the order of magnitude of $\delta\varepsilon_{FI}$ corrections is  10~meV or less, whereas the experimental uncertainties for chemical shifts in  most cases of common interest are larger.

%% file: tabl7.tex
\begin{table}
         \caption{The values of $R^{<>}_{kX}$ and $R^{<<}_{kX}$ for outer-core $np_{3/2}$ and $nd_{3/2}$ shells of the Pb atom when $k=2$.$^a$}
        \centering
	\begin{ruledtabular}
        \begin{tabular}{lcc}
         $X$ &  $R^{<>}_{kX}$, meV&     $R^{<<}_{kX}$, meV    \\
        \hline
        $2p_{3/2}$ & $15.5 $ & $155$                   \\
        $3p_{3/2}$ & $130$ & $243$ \\
        $3d_{3/2}$ & $120$ & $240$  \\
        \hline
        $4p_{3/2}$ &  $430$&  $163$ \\
        $4d_{3/2}$ &  $460$&  $163$\\
        \end{tabular} 
	\end{ruledtabular}
  \footnotetext[0]{The core radius $R_c=3.16$~a.u.\ is enlarged here (compared to the value $R_c=0.5$~a.u.\ used elsewhere in this paper) to cover inside the $4p_{3/2}$ and $4d_{3/2}$ shells.
}
  \label{tabl7}
\end{table}

%% file: app_c.tex
  \section{Derivation of contribution to the inner core transition energy from interaction of the TC electron with valence and outer core shells in the sudden transition approximation}
\label{app_c}

In Sec.~\ref{sec2} we show that the chemical shift of Ic transition energy in an atom bound in some compound ``$M$'' compared to the free neutral atom ``$A$'', $\chi_{FI}$, in the relativistic average configuration approximation is a difference of the corresponding mean values of the one-electron operator $\overline{\bm{\chi}_{FI}}$.

Let us denote the one-electron states belonging to shells $F$ and $I$ with fixed projection numbers of $m_f$ and $m_i$ as $\left | f \right \rangle$ and $\left| i \right \rangle$. The operator $\bm{\chi}_{fi}$ is a combination of Coulomb and exchange operators
\begin{equation}
        \bm{\chi}_{fi}= \bm{J}(f) - \bm{J}(i) - \bm{K}(f) +\bm{K}(i)\ ,
        \label{eq:chidef}
\end{equation}
the matrix elements of which are 
\[
J_{rs}(x) = \left\langle  rx|r_{12}^{-1}| sx \right\rangle,
\]
\[
K_{rs}(x) = \left\langle rx  |r_{12}^{-1}|xs  \right\rangle,\, x = i,f.
\]

 For the case of unconstrained averaging (over all the possible indices $m_i$ and $m_f$), the average operator $\overline{\chi_{FI}}$ is the sum of average operators $J$ and $K$:

\begin{equation}
        \bm{\overline{\chi_{FI}}} = \overline{\bm{J}}({F}) - \overline{\bm{J}}({I}) -
        \overline{\bm{K}}({F}) + \overline{\bm{K}}({I}). 
        \label{eq:endapp}
\end{equation}
        
Applying the Wigner-Eckart theorem, one can express the corresponding two-electron matrix elements as
follows (e.g., see Ref.~\cite{Grant:88} for details): 

\begin{equation}
        \begin{split}
	\left\langle rx |r_{12}^{-1}| sx \right\rangle =
        \sum\limits_{k}(-1)^{j_r-m_r+j_x-m_x+k}\langle rx ||V_{k}||sx\rangle\cdot\\
        \cdot \left ( \begin{array}{ccc} j_x& k& j_x \\ -m_x& 0& m_x \end{array}\right)\left ( \begin{array}{ccc} j_r& k& j_s \\ -m_r& 0& m_s \end{array}\right)\ ,
    \end{split}
\end{equation}

\begin{equation}
        \begin{split}
	\left\langle rx |r_{12}^{-1}| xs  \right\rangle
    =\sum\limits_{kq}(-1)^{j_r-m_r+j_x-m_x+k-q}\langle rx ||V_{k}||xs\rangle\cdot\\
    \cdot\left ( \begin{array}{ccc} j_x& k& j_r \\ m_x& -q& -m_r \end{array}\right)\left ( \begin{array}{ccc} j_x& k& j_s \\ -m_x& q& m_s \end{array}\right)\ ,
    \end{split}
	\label{eq:l3}
\end{equation}
where $m_x,\ m_r$, and $m_s$ are magnetic quantum numbers of the corresponding one-electron states; $\langle rx ||V_{k}||sx\rangle$ are the reduced matrix elements which are independent of the $m_x,\ m_r$, and $m_s$ indices. These matrix elements are proportional to the radial integrals $R_k(rxsx)$. In the case of Coulomb interaction between electrons, $V=r^{-1}_{12}$, and for $\alpha,\,\beta,\,\gamma,\,\delta \in$ W,\, Ic, the former integrals are equal to

\begin{equation}
\begin{array}{c}
	R_k(\alpha\beta\gamma\delta) = \int\limits_{0}^{R_c}\int\limits_{0}^{R_c}r_1^2dr_1\,r_2^2dr_2\ ,
	\rho^e_{\alpha\gamma}(r_1)\rho^e_{\beta\delta}(r_2)\frac{r_<^k}{r_>^{k+1}},\\ \\
	\rho^e_{\xi\zeta}(r) = p_\xi(r)p^*_\zeta(r) +
        q_\xi(r)q^*_\zeta(r)\mbox{, при $\xi,\zeta=\alpha,\beta,\gamma,\delta\ ,$}
\end{array}
	\label{eq:l4}
\end{equation}
In these expressions $p_\xi(r),\, q_\xi(r)$ are the large and small components of the 
one-electron state $\left | \xi \right \rangle$.

Let us first consider the direct Coulomb matrix elements $J_{rs}(x)$. Using the equality \cite{Varshalovich:88}

\begin{equation*}
        \sum\limits_m(-1)^{j-m}\left ( \begin{array}{ccc} j& k& j \\ -m& 0& m \end{array}\right)=\delta_{k0}\sqrt{2j+1},
        \label{eq:Vars}
\end{equation*}
\vbox{
we obtain

\begin{equation*} 
\begin{split}
        \overline{J}_{rs}({X})= \frac{1}{2j_x+1}\sum\limits_{i}\left\langle
        rx|r_{12}^{-1}|sx  \right\rangle = \\
        \langle rx ||V_{0}||sx\rangle
        \sqrt{\frac{2j_r+1}{2j_x+1}}\,\delta_{m_rm_s}\delta_{j_rj_s}\ .
\end{split}
        \label{eq:l5} 
\end{equation*}
}
Consider the exchange matrix elements $\left\langle rx |r_{12}^{-1}| xs \right\rangle$. Taking into account
\begin{equation*}
    \left ( \begin{array}{ccc} j_i& k& j_r \\ m_x& -q& -m_r \end{array}\right) \ne 0\ ,
\end{equation*}
 one can write 

\begin{equation}
        \begin{array}{c}
                m_x-q-m_r = 0, \\
                (-1)^{j_r-m_r+j_i-m_x+k-q} =(-1)^{1+j_r+j_i+k}.
        \end{array}
        \label{eq:exc1}
\end{equation}
\vbox{
Using the equalities \cite{Varshalovich:88}
\begin{equation*}
 \begin{array}{c} \left ( \begin{array}{ccc} j_r& j_s& j_x \\ m_r& m_s& m_x \end{array}\right){=}
        \left ( \begin{array}{ccc} j_r& j_s& j_x \\ -m_r& -m_s& -m_x \end{array}\right)(-1)^{j_a+j_s+j_x},\\ \\
        \sum\limits_{m_rm_s}\left ( \begin{array}{ccc} j_r& j_s& j_x \\ m_r& m_s& m_x \end{array}\right)\left ( \begin{array}{ccc} j_r& j_s& j_x' \\ m_r& m_s& m_x' \end{array}\right)
        = \frac{\delta_{m_xm_x'}\delta_{j_x j_x'}}{2j_x+1}, \end{array}
        \label{eq:exc2} \end{equation*}
}
we obtain

\begin{equation}
        \overline{K}_{rs}(x)=\frac{\delta_{m_r,m_s}\delta_{j_r,j_s}}{(2j_r+1)(2j_x+1)}\sum\limits_{k}\langle rx ||V_{k}||xs\rangle\ .        
        \label{eq:exc3}
\end{equation}
The final expression for the $\overline{\bm{\chi}_{FI}}$ matrix elements
   with independent averaging over $m_i$ and $m_f$
is

\begin{equation*}
\begin{split}
\overline{\chi_{FI}}^{rs}=\delta_{j_r,j_s}\delta_{m_r,m_s}(\overline{J}_{rs}({F})-\overline{J}_{rs}({I}) - \\
            -\overline{K}_{rs}({F})+\overline{K}_{rs}({I})).
\end{split}
\end{equation*}